\documentclass[runningheads]{llncs}
\usepackage{amsmath}
\usepackage[T1]{fontenc}
\usepackage{wasysym}
\usepackage{graphicx}
\usepackage{svg}
\usepackage{pgfplots}
\usepackage{subcaption}
\usepackage{threeparttable}
\usepackage{orcidlink}
\usepackage{booktabs}
\usepackage{multirow}
\usepackage{url}
\usepackage[backend=biber, style=numeric-comp, sorting=none, maxbibnames=99]{biblatex}
\usepackage{tikz}
\usetikzlibrary{positioning, shapes.geometric, arrows.meta, patterns}
\pgfplotsset{compat=1.18}
\title{Evaluating Differential Privacy Against Membership Inference in Federated Learning: Insights from the NIST Genomics Red Team Challenge}
\author{Gustavo de Carvalho Bertoli \inst{}\orcidlink{0000-0003-1940-8295}}
\authorrunning{G. de Carvalho Bertoli}
\titlerunning{Differential Privacy Against Membership Inference in Federated Learning}
\institute{\,
}

\begin{document}

\maketitle

\begin{abstract}
    While Federated Learning (FL) mitigates direct data exposure, the resulting trained models remain susceptible to membership inference attacks (MIAs). This paper presents an empirical evaluation of Differential Privacy (DP) as a defense mechanism against MIAs in FL, leveraging the environment of the 2025 NIST Genomics Privacy-Preserving Federated Learning (PPFL) Red Teaming Event. 
    To improve inference accuracy, we propose a stacking attack strategy that ensembles seven black-box estimators to train a meta-classifier on prediction probabilities and cross-entropy losses. 
    We evaluate this methodology against target models under three privacy configurations: an unprotected convolutional neural network (CNN, $\epsilon=\infty$), a low-privacy DP model ($\epsilon=200$), and a high-privacy DP model ($\epsilon=10$). 
    The attack outperforms all baselines in the No DP and Low Privacy settings and, critically, maintains measurable membership leakage at $\epsilon=200$ where a single-signal LiRA baseline collapses. Evaluated on an independent third-party benchmark, these results provide an empirical characterisation of how stacking-based inference degrades across calibrated DP tiers in FL.
\end{abstract}

\section{Introduction}
Federated Learning (FL) is a privacy-enhancing technology that enables machine learning (ML) tasks to be performed without exchanging raw data~\cite{fl}. In addition, FL enables collaboration across privacy-sensitive domains such as healthcare, security, and banking, where data sharing directly is legally or ethically constrained~\cite{healthcare, brain, crime, banking}.

Despite these benefits, FL introduces its own security and privacy risks, including malicious servers, poisoning attacks, backdoors, and membership inference attacks (MIAs)~\cite{survey-fl-sec-priv}.
MIAs are risky in sensitive domains. At the individual level, a successful MIA confirms whether a specific record was used for training — raising concerns for data subjects who did not consent to inclusion or who wish to verify compliance with data protection regulations. At the adversarial level, knowledge of training membership serves as reconnaissance: an attacker who knows which records a model was trained on can craft more targeted evasion or extraction attacks~\cite{carlini2022}. More broadly, MIAs serve as an auditing tool — they provide an empirical means of verifying whether a privacy-preserving mechanism, such as differential privacy, bounds membership leakage in practice.

To address privacy concerns, Privacy-Preserving Federated Learning (PPFL) extends FL with additional mechanisms to reduce what can be learned across all the FL participants or the final central model derived from them. A common approach is for FL clients to add calibrated noise to their local model updates before sharing them with the central server.

This paper uses the NIST Genomics PPFL Red Teaming Event as a controlled environment to empirically assess how well Differential Privacy (DP) protects FL models against MIAs. The attack exploits the tendency of machine learning models to exhibit different behaviour on training versus non-training samples, particularly in terms of prediction confidence and loss. This effect is amplified in scenarios where models are over-fitted or trained on limited or heterogeneous data distributions.

The methodology achieved the highest aggregate performance in the challenge's membership inference track, ranking first in the No DP ($\epsilon=\infty$) and Low Privacy ($\epsilon=200$) tiers. The contributions of this paper are threefold: (1)~we propose a black-box, stacking-based MIA that replaces the canonical shadow-model requirement with auxiliary-record-driven meta-learning, making it compatible with FL threat models where private training data is inaccessible; (2)~we provide an empirical characterisation of how this multi-signal attack degrades across three calibrated DP tiers, showing that residual leakage persists at $\epsilon=200$ even when single-signal baselines collapse; and (3)~we evaluate on an independent, third-party genomic benchmark, establishing a reproducible baseline for future MIA research on this publicly available dataset.

The remainder of this paper is organised as follows. Section~\ref{sec:problem_setting} describes the problem setting. Section~\ref{sec:methodology} presents the attack methodology. Section~\ref{sec:results} reports the empirical results and further discussions including the evaluation metric and the trade-off between privacy and utility, followed by a discussion of limitations in Section~\ref{sec:limitations}. Section~\ref{sec:related} discusses related work. Section~\ref{sec:conclusion} concludes.

\section{Problem Setting}\label{sec:problem_setting}

This work builds on the NIST Genomics Red Team Challenge~\footnote{2025 Red Teaming Event: \url{https://pages.nist.gov/genomics_ppfl/}}, an initiative that evaluated the effectiveness of DP as a defence against two attack types in FL: membership inference and reconstruction attacks. For genomic data, the challenge presents a primary scenario on FL for predicting seed coat colour in soybean plants, with the attributes of the records being gene variants. A second scenario uses dog breed prediction as a proxy for human genomic data. This work focuses on the soybean dataset and membership inference. The NIST red teaming datasets are available in a public repository~\footnote{NIST PPFL Dataset: \url{https://github.com/usnistgov/genomics_ppfl}}.

\begin{figure}[htb]
    \centering
    \includegraphics[width=\linewidth]{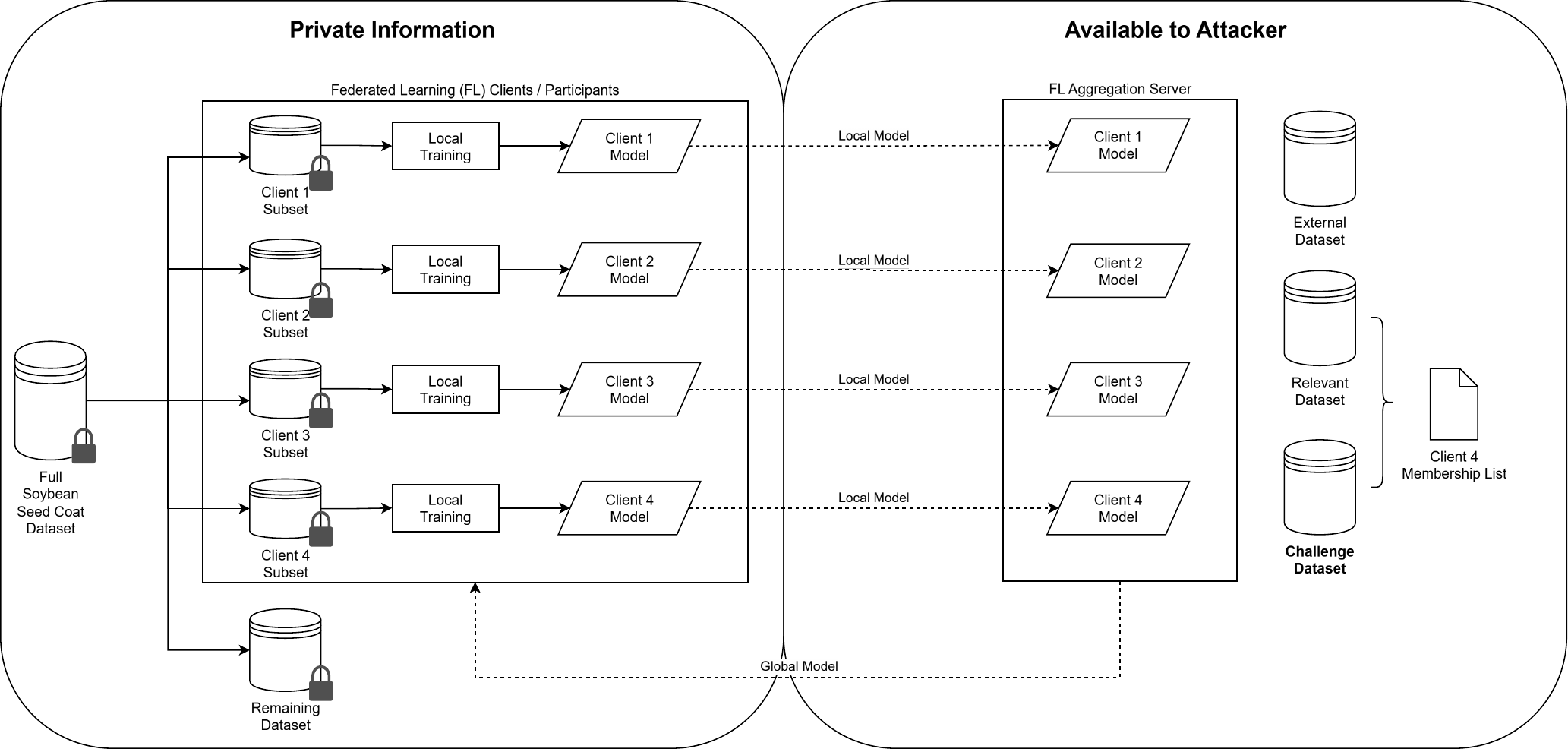}
    \caption{Problem setting. Clients train local models on private genomic data and share them with a central aggregator. The adversary queries client models and uses auxiliary data (relevant, external, challenge) to execute MIA.}
    \label{fig:problem-design}
\end{figure}

For this challenge, the soybean seed coat colour dataset features were reduced to the gene variants most relevant to the task, and the learning task is to classify records into one of four seed coat colour categories. The soybean seed coat dataset is split into five parts, four out of these five splits are used to train each of the four clients. The FL task in the challenge is restricted to $4$ clients, the remaining subset represents records that are not used by any of the clients during its training. This data split as well as the attacker available data are presented by Figure~\ref{fig:problem-design}.
Each client's model is evaluated across three privacy configurations: No DP (standard CNN, $\epsilon=\infty$), Low DP ($\epsilon=200$), and High DP ($\epsilon=10$). 

From the attacker's perspective, three datasets are available in addition to the trained models: the \textit{relevant dataset} represents complete records (with the correct target attribute) that might or might not be part of any of clients subsets, the \textit{external dataset} represents complete records (with the correct target attribute) that is guaranteed not to be part of any of clients subsets, and the \textit{challenge dataset} represents the blind evaluation set, for which the attacker must predict client membership. 

\subsection{Threat Model}\label{sec:threatmodel}

The red-teaming challenge proposes the untrusted central aggregator threat model. In this setting, federated learning (FL) clients collaboratively train local models on private genomic data and periodically share their models with the central server responsible for aggregation.

The adversary is the central server, which is assumed to be malicious. It has access to trained client models via black-box queries, i.e., the ability to submit arbitrary inputs and observe output predictions. Furthermore, as the central coordinator of the federated learning process, the adversary possesses full knowledge of the global training hyper-parameters, including the optimizer, learning rate, local epochs, and applied differential privacy constraints.

The adversary also has auxiliary datasets, provided as part of the challenge, as presented by Figure~\ref{fig:problem-design}. This capability is realistic: similar auxiliary data could be purchased from data brokers, consistent with the challenge design. The attack is black-box, relying only on the trained models without access to gradients, or internal parameters.

Additionally, the adversary does not have direct access to the original private datasets used by each client. However, this attack requires an additional assumption: one client colludes with the server. Consequently, the attacker knows which records in the \textit{relevant} and \textit{challenge} datasets belong to the colluding client. 

The adversary's goal is to carry out a membership inference attack (MIA), i.e., to determine whether a given data record was part of a specific client’s local training dataset. To mitigate this risk, clients may apply DP during training by injecting noise into gradient updates during local training. The challenge provides three privacy regimes. No Privacy (No DP): standard models trained without noise ($\epsilon = \infty$) for 100 epochs; Low Privacy (Low DP): $\epsilon = 200$, clipping at $2.0$, $100$ epochs; High Privacy (High DP): $\epsilon = 10$, clipping at $2.0$, $100$ epochs.

This threat model represents a realistic worst case: the central coordinator, colluding with one FL client, becomes the adversary.

\subsection{Dataset Dimensionality and Distribution}

As established in the threat model, the adversary relies on auxiliary datasets (\textit{relevant} and \textit{external}) to retrieve information for a successful attack. The dataset dimensions and distributions are identical across all three privacy regimes (No DP, Low DP, and High DP).

Each genomic record is represented by $125,767$ gene variants, forming a high-dimensional feature space. Table~\ref{tab:datasets} details the exact distribution of the auxiliary records available to the attacker for each target client, alongside the blind \textit{challenge} records that serve as the ultimate evaluation target. 

\begin{table}[htb]
\centering
\caption{Distribution of Auxiliary and Challenge Records per Target Client}
\label{tab:datasets}
\begin{tabular}{@{}l @{\hspace{4em}} c @{\hspace{4em}} c@{}}
\toprule
\textbf{Target} & \textbf{Relevant Records} & \textbf{External Records} \\ \midrule
Client 1 & 73 & 64 \\
Client 2 & 95 & 83 \\
Client 3 & 59 & 52 \\
Client 4 & 23 & 20 \\ \midrule
\textbf{Challenge Records} & \multicolumn{2}{c}{73 (evaluated collectively)} \\
\textbf{Feature Space} & \multicolumn{2}{c}{125,767 gene variants} \\ \bottomrule
\end{tabular}
\end{table}

This distribution highlights a vulnerability inherent to this federated learning task: the disparity between the number of available samples ($n$, a maximum of 95 relevant records for Client 2) and the vast feature space ($p = 125,767$). This high-dimensionality, low-sample regime ($n \ll p$) predisposes local models to severe over-fitting, which amplifies the membership signal exploited by MIAs.

\subsection{Target Model Architectures}

\begin{figure}[htb]
    \centering
    
    \begin{subfigure}[b]{\textwidth}
        \includegraphics[width=\textwidth]{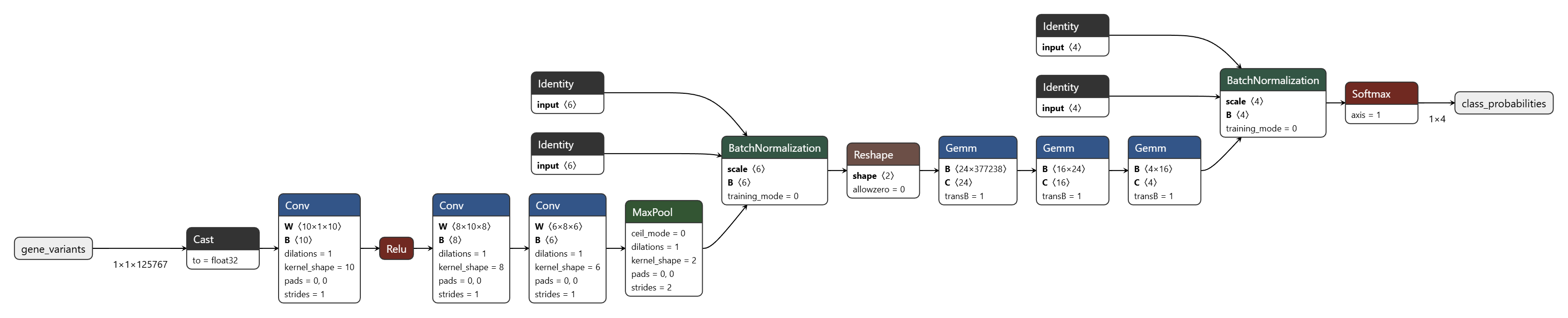}
        \caption{Non-DP CNN utilizing Batch Normalization.}
        \label{fig:cnn_arch}
    \end{subfigure}
    
    \vspace{1.5em}
    
    \begin{subfigure}[b]{\textwidth}
        \includegraphics[width=\textwidth]{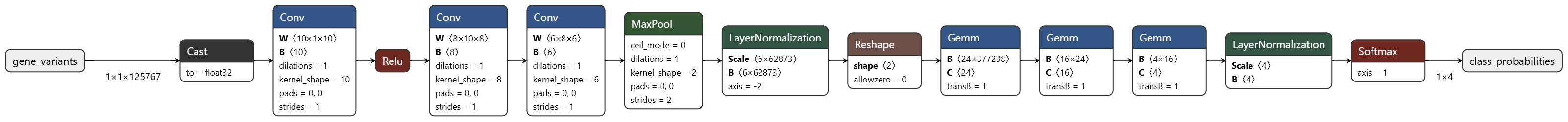}
        \caption{DP-enabled models utilizing Layer Normalization.}
        \label{fig:dpcnn_arch}
    \end{subfigure}
    
    \caption{Architectural comparison across privacy tiers. Batch Normalization (Non-DP CNN) is replaced by Layer Normalization in DP-enabled models to satisfy per-sample gradient clipping requirements.}
    \label{fig:architectures}
\end{figure}

The target models (Figure~\ref{fig:architectures}) share a 1D CNN architecture comprising three convolutional layers (kernel sizes 10, 8, 6), max-pooling, and a three-layer multi-layer perceptron (MLP) projecting to 4 output classes, with ReLU activations and Dropout during training.

The architecture differs between privacy tiers to accommodate the mathematical constraints of DP. As shown in Figure~\ref{fig:cnn_arch}, the unprotected baseline CNN utilizes standard Batch Normalization to stabilize training. However, because Batch Normalization computes statistics across multiple samples in a batch — inherently violating the strict per-sample independence required for gradient clipping in DP-SGD — the low-DP and high-DP models replace all batch normalization layers with Layer Normalization (Figure~\ref{fig:dpcnn_arch}). This architectural adaptation ensures that the sensitivity of each genomic record remains strictly bounded during private training.

\subsection{Differential Privacy (DP)}

Differential Privacy (DP) offers a mathematical standard for privacy by ensuring an output is insensitive to the inclusion or exclusion of any single individual's data \cite{dwork}. Formally, a randomized function $\mathcal{K}$ gives $\epsilon$-differential privacy if for all data sets $D_1$ and $D_2$ differing on at most one element, and all $S \subseteq \text{Range}(\mathcal{K})$, the following condition is met:

$$Pr[\mathcal{K}(D_1) \in S] \le \exp(\epsilon) \times Pr[\mathcal{K}(D_2) \in S]$$ 

The privacy parameter, $\epsilon$, formally bounds this insensitivity. A mechanism $\mathcal{K}$ satisfying this definition ensures that the removal of any single participant's data does not significantly change the probability of any output.

DP mechanisms calibrate noise to the $L_1$-sensitivity $\Delta f = \max_{D_1, D_2} ||f(D_1) - f(D_2)||_1$, ensuring output distributions shift by at most $\exp(\epsilon)$ across neighbouring datasets~\cite{dwork}. In practice, DP-SGD applies this framework to gradient descent via the Gaussian mechanism; throughout this paper, $\epsilon$ values are reported under $\delta = 1 \times 10^{-5}$ as configured in the challenge.

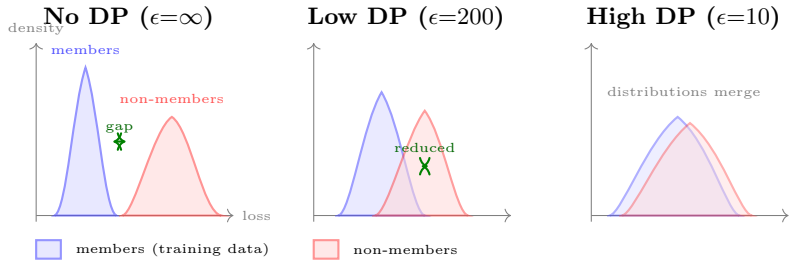
\begin{figure}[htb]
\centering
\begin{tikzpicture}[scale=0.82]

\tikzset{
  member/.style={draw=blue!70, fill=blue!15, thick},
  nonmember/.style={draw=red!70, fill=red!15, thick},
  axline/.style={->, thin, gray}
}

\foreach \xoff/\lab in {0/No DP ($\epsilon{=}\infty$), 4.5/Low DP ($\epsilon{=}200$), 9/High DP ($\epsilon{=}10$)} {
  \node[font=\small\bfseries] at (\xoff+1.5, 3.2) {\lab};
}

\begin{scope}[xshift=0cm]
  \draw[axline] (0,-0.1) -- (0,2.8) node[above,font=\tiny]{density};
  \draw[axline] (0,0) -- (3.2,0) node[right,font=\tiny]{loss};
  \filldraw[member, opacity=0.6]
    (0.3,0) .. controls (0.5,0.1) and (0.6,1.8) .. (0.8,2.4)
             .. controls (1.0,1.8) and (1.1,0.1) .. (1.3,0) -- cycle;
  \filldraw[nonmember, opacity=0.6]
    (1.4,0) .. controls (1.6,0.05) and (1.9,1.4) .. (2.2,1.6)
             .. controls (2.5,1.4) and (2.7,0.05) .. (3.0,0) -- cycle;
  \node[font=\tiny, blue!70] at (0.8,2.7) {members};
  \node[font=\tiny, red!70] at (2.2,1.9) {non-members};
  \draw[<->, thick, green!50!black] (1.25,1.2) -- (1.45,1.2)
    node[midway, above, font=\tiny, green!40!black]{gap};
\end{scope}

\begin{scope}[xshift=4.5cm]
  \draw[axline] (0,-0.1) -- (0,2.8);
  \draw[axline] (0,0) -- (3.2,0);
  \filldraw[member, opacity=0.55]
    (0.4,0) .. controls (0.6,0.05) and (0.8,1.5) .. (1.1,2.0)
             .. controls (1.4,1.5) and (1.6,0.05) .. (1.8,0) -- cycle;
  \filldraw[nonmember, opacity=0.55]
    (1.0,0) .. controls (1.2,0.05) and (1.5,1.3) .. (1.8,1.7)
             .. controls (2.1,1.3) and (2.3,0.05) .. (2.5,0) -- cycle;
  \draw[<->, thick, green!50!black] (1.75,0.8) -- (1.85,0.8)
    node[midway, above, font=\tiny, green!40!black]{};
  \node[font=\tiny, green!40!black] at (1.8,1.1) {reduced};
\end{scope}

\begin{scope}[xshift=9cm]
  \draw[axline] (0,-0.1) -- (0,2.8);
  \draw[axline] (0,0) -- (3.2,0);
  \filldraw[member, opacity=0.45]
    (0.3,0) .. controls (0.5,0.05) and (0.9,1.2) .. (1.4,1.6)
             .. controls (1.9,1.2) and (2.2,0.05) .. (2.4,0) -- cycle;
  \filldraw[nonmember, opacity=0.45]
    (0.5,0) .. controls (0.7,0.05) and (1.1,1.1) .. (1.6,1.5)
             .. controls (2.1,1.1) and (2.4,0.05) .. (2.6,0) -- cycle;
  \node[font=\tiny, gray] at (1.5,2.0) {distributions merge};
\end{scope}

\filldraw[member, opacity=0.6] (0,-0.7) rectangle (0.4,-0.4);
\node[font=\tiny, anchor=west] at (0.5,-0.55) {members (training data)};
\filldraw[nonmember, opacity=0.6] (4.5,-0.7) rectangle (4.9,-0.4);
\node[font=\tiny, anchor=west] at (5.0,-0.55) {non-members};

\end{tikzpicture}
\caption{Schematic effect of DP noise on member/non-member loss distributions. Without DP the distributions are well separated (left); as $\epsilon$ decreases, noise progressively closes the gap that MIAs exploit (right).}
\label{fig:dp_confidence_gap}
\end{figure}

The consequence for membership inference is direct. MIAs exploit the gap between the loss distributions of training members and non-members — a gap that exists because models tend to fit training data more closely than unseen data~\cite{yeom}. 
As $\epsilon$ decreases, DP noise flattens the model's per-sample confidence, progressively narrowing this gap until member and non-member records become indistinguishable. Figure~\ref{fig:dp_confidence_gap} illustrates this effect 
schematically across the three privacy tiers evaluated in this paper.

\section{Attack Methodology}\label{sec:methodology}
The challenge's black-box setting restricts the attacker to model output probabilities. Under this constraint, no single inference signal proves uniformly reliable: confidence-based signals degrade under DP noise, loss-based signals depend on the degree of over-fitting, and client heterogeneity means that what works for one client may fail for another. This motivates the stacking approach proposed in this paper: seven diverse base estimators each extract a partial membership signal, and a meta-classifier learns to combine them adaptively. The remainder of this section describes the pipeline. 

\subsection{Experimental Setup}
To evaluate the robustness of the client FL models against privacy threats, we utilized the Adversarial Robustness Toolbox (ART), specifically version 1.18.2 \cite{ART}. To ensure reproducibility of this adaptive attack evaluation, the source code for our methodology is available in a public repository\footnote{Repository: \url{https://github.com/gubertoli/nist-ppfl-mia}}. 

Across all 4 clients in the 3 privacy tiers, the client models share a common baseline configuration, trained for 100 epochs with a learning rate of $0.003$, a weight decay of $1 \times 10^{-4}$, a batch divisor of 15, and a test fraction of $0.3$ for the 4-class classification task (soy bean seed colour). The standard, non-private CNN utilizes the Adamax optimizer. Both DP models use SGD with DP-SGD-specific parameters: gradient clipping at norm $2.0$ and $\delta$ of $1 \times 10^{-5}$.

\subsection{Ensemble-based Membership Inference Attack (MIA)}\label{sec:ensemble_mia}
The black-box constraint of this threat model rules out the canonical shadow model strategy. Carlini et al.~\cite{carlini2022} establish that the most powerful MIA formulations require constructing paired IN/OUT shadow models on data drawn from the same distribution as the target's training set, so that per-record loss distributions can be estimated. In the untrusted central aggregator setting, this is not feasible: the adversary has no access to clients' private training partitions, and the available auxiliary pool is far too small (a maximum of 95 relevant records for any client) to train statistically reliable shadow models. 

This constraint shapes the feature design of the meta-dataset. Yeom et al.~\cite{yeom} showed that a simple loss threshold — classifying a record as a member if its loss falls below the average training loss — is a surprisingly effective baseline that follows directly from the definition of over-fitting. Salem et al.~\cite{Salem2019-qc} extended this insight, demonstrating that shadow-model-free attacks relying on confidence scores alone remain effective under relaxed adversarial assumptions. 

The meta-classifier combines both signals: the cross-entropy loss captures the over-fitting signal directly, while the seven ART base estimators contribute diverse probability-based views of the same signal — fusing the loss-threshold intuition of Yeom et al. with the confidence-score approach of Salem et al., without requiring shadow models or white-box access. Notably, Salem et al.~\cite{Salem2019-qc} also propose stacking as a defence; here the same paradigm is repurposed offensively.

We employ stacking~\cite{WOLPERT1992241} to aggregate these heterogeneous signals through a learned meta-model. The remainder of this section details the three components of the pipeline: the base attack models (Section~\ref{sec:base_attacks}), the meta-feature construction (Section~\ref{sec:meta_features}), and the meta-model with domain adaptation (Section~\ref{sec:meta_model}).

\subsubsection{Base Attack Models}\label{sec:base_attacks}

Rather than relying on a single inference algorithm, we utilized the ART to train a diverse set of seven black-box MIA models. The target FL models (both standard CNN and DP-CNN configurations) are queried as frozen black-box oracles. The base attack algorithms trained on these outputs included: Neural Network (NN), Random Forest (RF), Decision Tree (DT), Gradient Boosting (GB), K-Nearest Neighbors (KNN), Support Vector Machine (SVM), and Logistic Regression (LR). 

Following black-box MIA formulations, the ART estimators are instantiated as class-dependent attack models. During training, the target model's continuous 4-class output probabilities $f(X)$ are concatenated with the record's true task label $y$ (one-hot encoded). Because the attacker lacks perfect membership ground truth for the target clients, the base estimators are trained using a noisy proxy: the provided \textit{relevant} records serve as the assumed member class, while the \textit{external} records serve as confirmed non-members.

This labelling is deliberately imprecise. The \textit{relevant} dataset contains a mixture of true members and non-members; for Client~4, where ground truth is available, 13 of 23 relevant records are actual members, 
yielding a label noise rate of approximately 43\%. The standard MIA assumption -- that the adversary can cleanly label auxiliary data drawn from the target  distribution~\cite{reza} -- does not hold here. Despite this, the base estimators remain useful because they are not used for final membership decisions. Their role is to produce continuous probability scores that become \textit{features} for the meta-classifier. Even when a base estimator's decision boundary is shifted by label noise, the relative ordering of its output probabilities can still carry membership signal. The meta-classifier, trained on Client~4's clean labels, then learns to calibrate these noisy scores against ground truth. This two-stage design -- noisy base training followed by clean meta-learning -- makes the pipeline tolerant of imperfect auxiliary labelling, a condition likely to arise in realistic FL adversarial scenarios where the attacker cannot verify membership of auxiliary records.

Once trained, the seven base attack models are applied to the \textit{challenge} records, producing continuous membership probabilities that feed the meta-classifier.

\subsubsection{Meta-Feature Engineering}\label{sec:meta_features}

For each target record $X$, we query all seven trained ART estimators for their continuous membership probabilities, and compute the target model's cross-entropy loss $\mathcal{L}_{CE}$. 
This extraction results in an 8-dimensional feature vector for each sample:

$$X_{meta} = [p_{NN}, p_{RF}, p_{DT}, p_{GB}, p_{KNN}, p_{SVM}, p_{LR}, \mathcal{L}_{CE}]$$

where $p_{model}$ represents the probability of membership inferred by each respective base attack model. Because the adversary only possesses leaked ground-truth membership for one specific client (Client 4), this client serves as the proxy for meta-learning. Consequently, the target variable $y$ in the meta-dataset was constructed by setting $y=1$ for the known members within Client 4's \textit{relevant} and \textit{challenge} records, and $y=0$ for all confirmed non-members (comprising the remainder of the \textit{relevant}, \textit{challenge} records and the entirety of the \textit{external} records). Since Client 4's challenge membership is fully known to the attacker (Section~\ref{sec:threatmodel}), those records are directly assigned to Client 4 without passing through the meta-classifier; only the remaining challenge records require inference.

\subsubsection{Meta-Model Optimization and Domain Adaptation}\label{sec:meta_model}

We select eXtreme Gradient Boosting (XGBoost) as the final meta-classifier~\cite{xgboost}, given its established performance on tabular features with mixed scales. However, directly applying Client 4's trained meta-model to evaluate Clients 1, 2, and 3 risks distributional shift, as each local model possesses distinct confidence landscapes. To address this without requiring positive ground-truth labels for the remaining clients, we introduce a partial-fit domain adaptation strategy. For each subsequent target client, the base XGBoost model is incrementally updated (fine-tuned) using that client's \textit{external} records (guaranteed non-members, $y=0$). 

Following this adaptation, we apply a two-dimensional decision rule to assign challenge records. The original challenge submission resolved conflicts through visual inspection of the per-client probability distributions; this paper formalises that judgment as a reproducible criterion. 

The first condition (column) suppresses per-client noise: $P(X,c)>P^{(c)}_{thr}$, where $P^{(c)}_{thr}$ is the 55th percentile of the meta-classifier's output probabilities for client $c$ across all challenge records. This percentile was selected empirically on Client 4's labelled data as the lowest threshold that filters spurious non-member assignments while retaining members. The second condition (row) ensures the candidate client dominates the others for that record: $P(X,c)>\lambda\,\bar{P}(X)$, where $\bar{P}(X)=\frac{1}{3}\sum_{c'}P(X,c')$ and $\lambda=1.5$; this multiplier requires a client's score to exceed the cross-client average by 50\%, suppressing ambiguous assignments. Record $X$ is assigned to $c^*=\arg\max_c\,P(X,c)$ only when both conditions hold simultaneously; otherwise it is predicted as non-member. The column threshold adapts naturally to the privacy regime: under High DP ($\epsilon=10$) the diffuse probability landscape raises $P^{(c)}_{thr}\approx0.025$, while under Low DP ($\epsilon=200$) noise suppresses the overall confidence scale, lowering it to $\approx 0.001$.

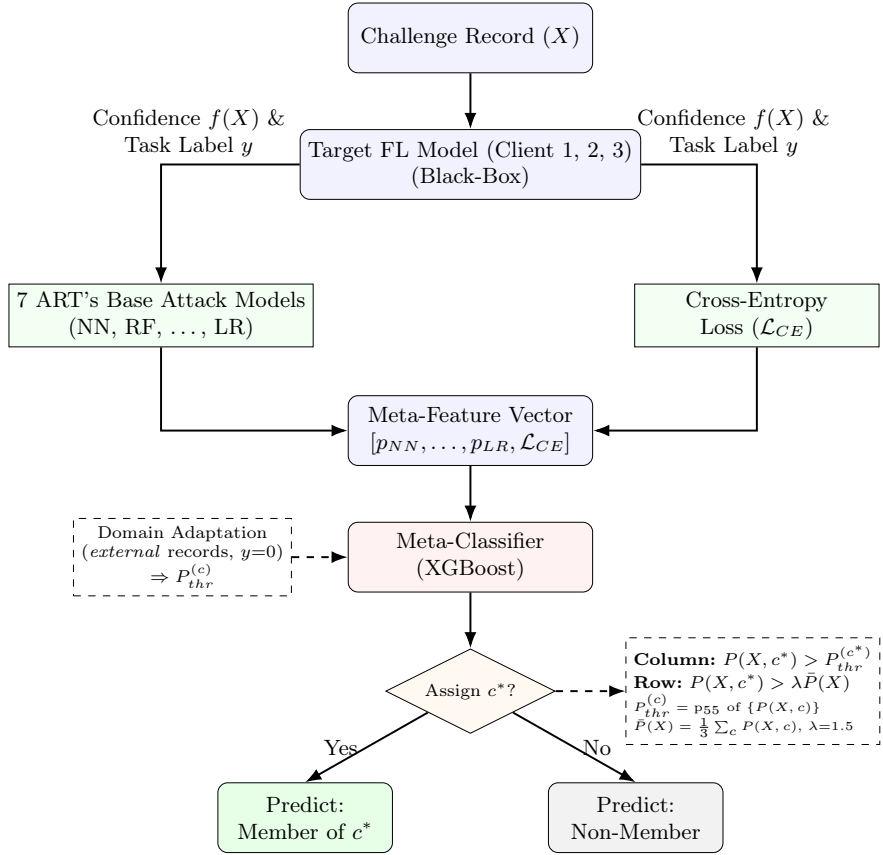
\begin{figure}[htbp]
    \centering
    \resizebox{0.95\linewidth}{!}{%
    \begin{tikzpicture}[
        node distance=1cm and 0.5cm,
        box/.style={draw, rectangle, align=center, minimum width=3.5cm, minimum height=1cm, rounded corners, fill=blue!5},
        estimator/.style={draw, rectangle, align=center, minimum width=3.5cm, minimum height=0.8cm, fill=green!5},
        diamond_node/.style={draw, diamond, aspect=2, align=center, fill=orange!5},
        arrow/.style={-Latex, thick}
    ]
    
    \node[box] (input) {Challenge Record ($X$)};
    \node[box, below=0.8cm of input] (target) {Target FL Model (Client 1, 2, 3)\\(Black-Box)};
    
    \node[estimator, below left=1.2cm and -0.2cm of target] (art) {7 ART's Base Attack Models\\(NN, RF, \dots, LR)};
    \node[estimator, below right=1.2cm and -0.1cm of target] (loss) {Cross-Entropy\\Loss ($\mathcal{L}_{CE}$)};
    
    \node[box, below=2.8cm of target] (vector) {Meta-Feature Vector\\$[p_{NN}, \dots, p_{LR}, \mathcal{L}_{CE}]$};
    
    \node[box, below=0.8cm of vector, fill=red!5] (xgboost) {Meta-Classifier\\ (XGBoost)};
    
    \node[draw, dashed, align=center, left=0.8cm of xgboost, font=\scriptsize] (adapt)
    {Domain Adaptation\\(\textit{external} records, $y{=}0$)\\$\Rightarrow P^{(c)}_{thr}$};

    \node[diamond_node, below=0.8cm of xgboost, align=center, font=\scriptsize]
        (thresh) {Assign $c^*$?};
    
    \node[draw, rectangle, dashed, right=1cm of thresh, align=left, font=\scriptsize]
        (rule) {%
          \textbf{Column:} $P(X,c^*) > P^{(c^*)}_{thr}$\\
          \textbf{Row:} $P(X,c^*) > \lambda\bar{P}(X)$\\
          {\tiny $P^{(c)}_{thr}=\text{p}_{55}$ of $\{P(X,c)\}$}\\
          {\tiny $\bar{P}(X)=\frac{1}{3}\sum_{c}P(X,c)$, $\lambda{=}1.5$}};
    
    \draw[arrow, dashed] (thresh) -- (rule);
    
    \node[box, below left=1cm and 0.5cm of thresh, fill=green!10, minimum width=2.5cm] (member) {Predict:\\Member of $c^*$};
    \node[box, below right=1cm and 0.5cm of thresh, fill=gray!10, minimum width=2.5cm] (nonmember) {Predict:\\Non-Member};

    \draw[arrow] (input) -- (target);
    
    \draw[arrow] (target) -| node[pos=0.4, above, align=center, font=\small] {Confidence $f(X)$ \&\\ Task Label $y$} (art.north);
    \draw[arrow] (target) -| node[pos=0.4, above, align=center, font=\small] {Confidence $f(X)$ \& \\Task Label $y$} (loss.north);
    
    \draw[arrow] (art) |- (vector);
    \draw[arrow] (loss) |- (vector);
    
    \draw[arrow] (vector) -- (xgboost);
    \draw[arrow, dashed] (adapt) -- (xgboost);
    
    \draw[arrow] (xgboost) -- (thresh);
    
    \draw[arrow] (thresh.south west) -- node[left, pos=0.5] {Yes} (member.north);
    \draw[arrow] (thresh.south east) -- node[right, pos=0.5] {No} (nonmember.north);
    
    \end{tikzpicture}%
    }
    \vspace{0.5em}
    \caption{Ensemble MIA inference pipeline. Solid arrows show the forward pass of a challenge record $X$ through the target FL model, seven ART base estimators, and the XGBoost meta-classifier. The dashed path denotes offline domain adaptation on known non-members, yielding the per-client noise floor $P^{(c)}_{thr}$; assignment requires both column ($P>P^{(c)}_{thr}$) and row ($P>1.5\,\bar{P}(X)$) conditions.}
    \label{fig:ensemble_arch}
\end{figure}

\section{Empirical Results}\label{sec:results}

The attack was evaluated in two phases: local validation on Client 4's ground-truth data, and a blind standardised evaluation via the 2025 NIST Genomics PPFL Red Teaming Event platform.

During the local validation on Client 4, the meta-model achieved $90.8\%$ mean cross-validation accuracy ($\sigma \pm 0.094$) using repeated 10-fold cross-validation (3 repeats), indicating that the stacked feature set captures a membership signal on the labelled subset. Table~\ref{tab:leaderboard} shows the controlled red-teaming evaluation, which provides a standardised, third-party measure of attack effectiveness. Our approach reports the highest accuracy for the No DP and Low Privacy settings. Breaking down the results across the three privacy tiers reveals a trade-off: the attack achieved a score of $53.42\%$ accuracy against the no-DP CNN, which is nearly double the accuracy of the Baseline 1 ($26.03\%$), suggesting the meta-classifier extracts richer membership signals than single-model baselines. 

Furthermore, given the fixed evaluation size of $n=73$ \textit{challenge} records, a two-proportion z-test confirms that our ensemble's performance in the No DP tier ($53.42\%$) represents a statistically significant improvement over the closest baseline 2 ($30.14\%$, $z=2.84$, $p < 0.01$). This confirms the meta-classifier is extracting genuine membership signals rather than exploiting random variance. Conversely, under the Low DP ($\epsilon=200$) tier, the injected noise introduces sufficient variance to render the margin between our attack ($38.36\%$) and the runner-up ($28.77\%$) statistically indistinguishable ($z=1.44$, $p > 0.05$). This validates the defense: even a low DP budget disrupts the stable confidence signals required to reliably isolate members.

\begin{table}[htb]
\centering
\caption{MIA methodologies on the Challenge Set (third-party evaluation)}
\label{tab:leaderboard}
\begin{tabular}{@{}c @{\hspace{2em}} c @{\hspace{2em}} c @{\hspace{2em}} c @{\hspace{2em}}}
\toprule
\multirow{2}{*}{\textbf{ }} & \multicolumn{3}{c}{\textbf{Accuracy on Challenge Records (\%)}} \\ \cmidrule(l){2-4} 
\multicolumn{1}{l}{} & No DP & \begin{tabular}[c]{@{}c@{}}Low Privacy\\ ($\epsilon=200$)\end{tabular} & \begin{tabular}[c]{@{}c@{}}High Privacy\\ ($\epsilon=10$)\end{tabular} \\ \midrule
This work & 53.42 & 38.36 & 24.66 \\
Baseline 1 & 26.03 & 28.77 & 26.03 \\
Baseline 2\footnotemark & 30.14 & 20.55 & 26.03 \\
Baseline 3 & 27.4 & 26.03 & 20.55 \\
Baseline 4 & 26.03 & 19.18 & 27.4 \\
Baseline 5 & 17.81 & 23.29 & 13.7 \\
Baseline 6 & 10.96 & 17.81 & 20.55 \\
Baseline 7 & 5.48 & 4.11 & 8.22 \\ \bottomrule
\end{tabular}
\end{table}
\footnotetext{The Baseline 2, also developed by the author, is an oversimplification of the LiRA~\cite{carlini2022}. Instead of training extensive shadow models to estimate per-example hardness as in canonical LiRA, this lightweight adaptation leveraged the provided \textit{external records} to establish a baseline out-of-distribution (non-member) profile for each client model. Specifically, the attack computed the mean cross-entropy loss and mean prediction confidence across the external dataset. A \textit{challenge} record was then inferred as a member if its evaluated loss was strictly lower than the mean external loss and its confidence exceeded the mean external confidence, ultimately assigning the record to the client model that yielded the highest confidence score. This heuristic mirrors the confidence-based attack of Salem et al.~\cite{Salem2019-qc}. }

\begin{figure}[htb]
\centering
\begin{tikzpicture}
\begin{axis}[
    ybar,
    bar width=5pt,
    width=\linewidth,
    height=5.5cm,
    ylabel={Accuracy (\%)},
    symbolic x coords={No DP, Low DP ($\epsilon{=}200$), High DP ($\epsilon{=}10$)},
    xtick=data,
    xticklabel style={font=\small},
    ymin=0, ymax=65,
    ytick={0,10,20,30,40,50,60},
    legend style={at={(1,1)}, anchor=north east,
                  font=\tiny, legend columns=2},
    legend image post style={draw=none},
    legend cell align=left,
    pattern=north east lines,
    nodes near coords,
    nodes near coords style={font=\tiny, rotate=90, anchor=west},
    every node near coord/.append style={yshift=2pt},
    enlarge x limits=0.25,
    grid=major,
    grid style={dashed, gray!40},
]

\addplot[fill=blue!70, draw=blue!90, bar shift=-10pt, forget plot] coordinates {
    (No DP, 53.42)
    (Low DP ($\epsilon{=}200$), 38.36)
    (High DP ($\epsilon{=}10$), 24.66)
};
\addlegendimage{area legend, fill=blue!70, draw=blue!90}
\addlegendentry{This work}

\addplot[fill=gray!50, draw=gray!70, bar shift=-3.5pt, forget plot] coordinates {
    (No DP, 30.14)
    (Low DP ($\epsilon{=}200$), 28.77)
    (High DP ($\epsilon{=}10$), 27.4)
};
\addlegendimage{area legend, fill=gray!50, draw=gray!70}
\addlegendentry{Best baseline}

\addplot[fill=gray!30, draw=gray!50, bar shift=3.5pt, forget plot] coordinates {
    (No DP, 26.03)
    (Low DP ($\epsilon{=}200$), 20.55)
    (High DP ($\epsilon{=}10$), 20.55)
};
\addlegendimage{area legend, fill=gray!30, draw=gray!50}
\addlegendentry{Median baseline}

\addplot[fill=gray!15, draw=gray!40, bar shift=10pt, forget plot] coordinates {
    (No DP, 26.03)
    (Low DP ($\epsilon{=}200$), 26.03)
    (High DP ($\epsilon{=}10$), 26.03)
};
\addlegendimage{area legend, fill=gray!15, draw=gray!40}
\addlegendentry{Random baseline ($\approx$26\%)}

\end{axis}
\end{tikzpicture}
\caption{Attack accuracy across privacy tiers. Under High Privacy ($\epsilon=10$), several baselines score comparably to this work, consistent with the distribution-merging effect shown in Figure~\ref{fig:dp_confidence_gap}. Random assignment ($\approx$26\%) provides a lower bound.}
\label{fig:accuracy_bars}
\end{figure}
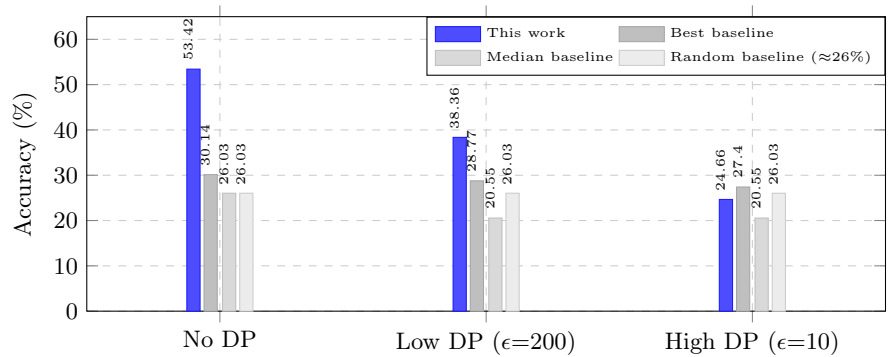

As expected, there is a monotonic degradation in attack accuracy as privacy constraints tighten (Figure~\ref{fig:dp_confidence_gap}). For our approach, success rates dropped from 53.42\% (No DP) to 38.36\% (Low Privacy, $\epsilon=200$) and finally to 24.66\% (High Privacy, $\epsilon=10$). This represents a total accuracy reduction of 53.8\% when moving from a non-private to a high-privacy setting.

In the High-DP ($\epsilon=10$) tier, the picture changes (Figure~\ref{fig:accuracy_bars}). The performance gap between submissions narrows noticeably — the top five baselines cluster between 20\% and 27\% — and several baselines marginally outperform this work (24.66\%). This convergence is consistent with the schematic in Figure~\ref{fig:dp_confidence_gap}: when distributions nearly merge, no method can reliably extract a signal, and results approach the random assignment floor of approximately 26\%. The ensemble's advantage, which derives from aggregating multidimensional signals, diminishes precisely when those signals become indistinguishable from noise.

\subsection{True-Positive Rate (TPR) at low False-Positive Rates}
As argued by \cite{carlini2022}, average-case accuracy provides an incomplete measure of privacy risk; attacks are better evaluated using the true-positive rate (TPR) at low false-positive rates (FPR). Because the blind challenge dataset does not provide public ground truth for all clients, we restrict this evaluation to Client 4, for which complete membership labels are available. Consequently, this analysis is isolated to a traditional ML setting rather than a full FL scenario, while preserving the identical dataset and model architectures to compare our ensemble proposal against the LiRA~\cite{carlini2022}.

We evaluate TPR at strict FPR boundaries ($\le 1\%$ and $\le 3\%$). For the ensemble approach, the meta-classifier is evaluated via 5-fold stratified cross-validation on Client 4's 116 labelled records; the out-of-fold membership probabilities $P(X)$ serve as the scoring metric for the ROC analysis, ensuring scores are not inflated by in-sample evaluation. For the comparative baseline, we utilize a variance-based formulation of LiRA. Unlike the simplified binary threshold used for Baseline 2 in the challenge, this implementation more closely approximates the original LiRA methodology by modelling the continuous out-of-distribution cross-entropy ($\mathcal{L}_{CE}$) loss. We estimate this variance directly from the external datasets rather than training canonical shadow models. Evaluating exactly 116 records for Client 4 (13 members and 103 non-members) provides a minimum measurable FPR step of 0.97\% (1 mistake out of 103).

The results (Table \ref{tab:mia_tpr_fpr}) highlight distinct adversary behaviours across the three privacy regimes. For the unprotected baseline, the LiRA approach yields a higher TPR (92.31\%) than the ensemble proposed (46.15\%) at the strict $\le 1.00\%$ FPR threshold, suggesting that normalized cross-entropy loss provides a highly separable signal for the most vulnerable outliers. However, at the slightly relaxed $\le 3.00\%$ boundary, both methods reach a 100\% TPR, indicating that the undefended CNN memorizes the training data sufficiently to allow complete extraction of the targeted members. 

Under Low DP ($\epsilon=200$), DP noise alters the attack dynamics: the injected noise masks the raw loss variance, severely reducing the LiRA baseline's TPR to 0.00\% at the 1.00\% FPR boundary and 7.69\% at the 3.00\% boundary. In contrast, the ensemble maintains a 30.77\% TPR at the 1.00\% FPR boundary and 38.46\% at the 3.00\% boundary. This demonstrates that while single-metric thresholds are highly sensitive to loose DP bounds, aggregating multidimensional signals enables the meta-classifier to capture residual membership leakage. 

The divergence between the two adversaries under Low DP warrants explanation. LiRA relies on a single continuous signal ($\mathcal{L}_{CE}$); when DP noise shifts and broadens the loss distributions of members and non-members, a univariate threshold rapidly loses separability. The ensemble, by contrast, operates on an 8-dimensional feature space. Although DP noise degrades each dimension, the noise originates in the target model's training process and therefore affects all base estimators through the same perturbed confidence outputs. However, the \textit{mapping} from perturbed confidence to membership probability differs across estimator families: tree-based estimators (RF, GB, DT) partition the probability space along axis-aligned splits, distance-based estimators (KNN) exploit local neighbourhood density, and parametric estimators (LR, SVM, NN) learn global decision surfaces. This architectural diversity preserves partial decorrelation among base predictions even under shared noise, allowing XGBoost to recover a composite membership signal from individually degraded features. The effect is analogous to the classical result in ensemble learning: stacking improves performance precisely when base learners make partially uncorrelated errors~\cite{WOLPERT1992241}.

\begin{table}[htb]
\centering
\caption{TPR at fixed FPR thresholds on Client 4 ($N_{IN}=13$, $N_{OUT}=103$).}
\label{tab:mia_tpr_fpr}
\begin{tabular}{@{\hspace{1em}}lc @{\hspace{2em}} c @{\hspace{2em}} c}
\toprule
\textbf{Privacy Tier} & \textbf{Target FPR} & \textbf{This work} & \textbf{LiRA} \\
\midrule
\multirow{2}{*}{No DP (CNN)} & $\le 1.00\%$ & 46.15\% & \textbf{92.31\%} \\
 & $\le 3.00\%$ & \textbf{100.00\%} & \textbf{100.00\%} \\
\midrule
\multirow{2}{*}{Low DP ($\epsilon=200$)} & $\le 1.00\%$ & \textbf{30.77\%} & 0.00\% \\
 & $\le 3.00\%$ & \textbf{38.46\%} & 7.69\% \\
\midrule
\multirow{2}{*}{High DP ($\epsilon=10$)} & $\le 1.00\%$ & 0.00\% & 0.00\% \\
 & $\le 3.00\%$ & 0.00\% & 0.00\% \\
\bottomrule
\end{tabular}
\end{table}

Under the strict $\epsilon=10$ constraint, both adversaries fail to identify members at low FPRs (0.00\% TPR). The calibrated noise effectively flattens the confidence distributions, preventing the isolation of high-confidence records and empirically validating the theoretical protections of DP-SGD.

As noted in Section~\ref{sec:methodology}, canonical LiRA is incompatible with this FL threat model due to the absence of shadow models. Baseline 2 approximates the OUT-distribution using the guaranteed non-member \textit{external} records instead.

\subsection{Privacy vs. Utility Trade-off}\label{sec:privutiltradeoff}
The three privacy tiers provide a natural axis for measuring how DP noise affects attack success. Against the unprotected CNN baseline (no noise, 100 training epochs), the ensemble achieved its highest score of 53.42\%. While this may appear modest in absolute terms, it is consistent with Carlini et al.~\cite{carlini2022}, who report that well-calibrated MIAs typically plateau between 53\% and 60\% accuracy on standard datasets, and it is more than double the runner-up baseline (Table~\ref{tab:leaderboard}).

Introducing DP noise at $\epsilon=200$ — a permissive budget that adds relatively little perturbation — reduced attack accuracy to 38.36\%, a drop of approximately 28\% relative to the unprotected baseline. Even at this loose privacy budget, the noise is sufficient to disrupt the confidence signals the meta-classifier relies on. Tightening the constraint to $\epsilon=10$ reduced accuracy further. The monotonic decline across tiers is consistent with DP theory — as $\epsilon$ decreases, local models generalise more evenly across training and non-training samples, narrowing the confidence gap the attacker exploits. Yet the ensemble remained effective under No DP and Low Privacy. Under High Privacy, accuracy converges toward the random floor, suggesting $\epsilon=10$ largely closes the exploitable gap.

The above captures the attacker's perspective. To complete the picture, the trade-off requires quantifying what the defender sacrifices in model utility when tightening DP constraints. The challenge fixes local training at 100 epochs, a regime that induces substantial over-fitting and is therefore artificially favourable to MIA. Real FL deployments typically run far fewer local epochs per round to limit communication overhead and client drift. 

\begin{figure}[htb]
    \centering
    \includegraphics[width=\linewidth]{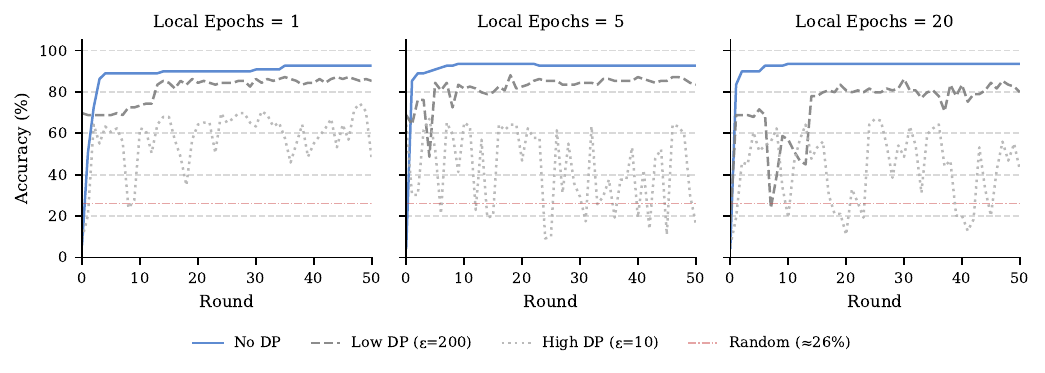}
    \caption{Per-round classification accuracy over 50 FL rounds across three privacy tiers and local epoch settings. The red dashed line marks the 4-class random assignment floor ($\approx$25\%).}
    \label{fig:fl_convergence}
\end{figure}

We simulate the FL pipeline using Flower~\cite{flower} across the three tiers with E $\in \{1, 5, 20\}$ local epochs over 50 rounds, using 80\% of labelled population data across 4 clients (20\% held out for evaluation). Figure~\ref{fig:fl_convergence} shows the per-round classification accuracy of the FL simulation across the three privacy tiers and local epoch settings. 

No DP converges reliably regardless of epoch count, reaching around 93\% by round 50, confirming that the non-private CNN learns effectively regardless of local computation budget. Low DP ($\epsilon=200$) retains useful utility — 83–85\% at E=1 and E=5 — though E=20 degrades to 79.8\%, suggesting that more local epochs amplify gradient clipping interference under DP-SGD. High DP ($\epsilon=10$) fails to converge in any epoch setting: round-50 accuracy ranges from 16\% to 47\% across the three configurations, remaining near or below the random assignment floor. This is the more consequential finding for the privacy-utility trade-off. The challenge's 100-epoch regime artificially favours the attacker by inducing over-fitting; under realistic FL conditions, $\epsilon=10$ does not merely reduce utility — it largely eliminates it. A defender forced to choose between a model that leaks membership under $\epsilon=200$ and one that fails to learn under $\epsilon=10$ faces a difficult operating constraint, and the gap between these two tiers is not bridgeable by tuning epochs alone.

Together, these two perspectives — degraded attack accuracy on one side, degraded model utility on the other — characterise the practical operating range of DP in this FL genomic setting.

\section{Limitations}\label{sec:limitations}
Several limitations bound the scope of these findings. First, the TPR evaluation relies on Client~4's 13 members and 103 non-members. At this scale, each correctly identified member shifts the TPR by approximately 7.7 percentage points, so the reported rates (e.g., 30.77\% at 1\% FPR under $\epsilon=200$) should be interpreted as evidence of residual leakage rather than precise point estimates. 

Second, because Client~4's membership is fully known, the TPR comparison against LiRA operates on a single client model in a traditional ML setting rather than across the FL setup; FL-specific dynamics such as aggregation artefacts are not captured in this analysis. 

Third, the results are specific to the NIST soybean genomics benchmark, where extreme dimensionality ($p = 125{,}767 \gg n$) amplifies over-fitting and may make membership signals more exploitable than in lower-dimensional domains. 

Generalisation to other FL applications, medical imaging or natural language processing, remains an open question. Finally, the Flower-based utility simulation (Section~\ref{sec:privutiltradeoff}) uses a balanced, IID data split that is more favourable than realistic FL partitions; utility degradation under non-IID conditions is likely to be more pronounced, further widening the privacy-utility gap identified in this work.

\section{Related Work}\label{sec:related}

Membership inference attacks (MIAs) were introduced by Shokri et al.~\cite{reza}, who showed that shadow-model-based adversaries can determine whether specific records were part of a model's training data. Subsequent work demonstrated that simpler signals — prediction confidence~\cite{Salem2019-qc} or loss~\cite{yeom} — are often sufficient, highlighting over-fitting as a primary source of leakage. However, these single-signal approaches have not been systematically evaluated against DP-protected models in FL settings.

In FL, privacy risks are further amplified due to heterogeneous client data and repeated model sharing. Prior work has shown that MIAs can be effective in FL settings, particularly under strong adversarial assumptions such as white-box access or gradient visibility~\cite{nasr,melis}. In contrast, more practical black-box attacks where only model outputs are accessible remain less explored.

DP-SGD~\cite{abadi}, which applies the DP framework~\cite{dwork} to gradient descent, has become the standard approach for training private models. While DP mathematically bounds membership leakage, empirical audits of these bounds often rely on weak or single-signal black-box attacks (e.g., simple loss thresholding)~\cite{yeom,jayaraman}. As Carlini et al.~\cite{carlini2022} argue, evaluating defences against non-adaptive adversaries is necessary but insufficient, as it risks underestimating worst-case privacy leakage.

Closest to our approach, Ullah et al.~\cite{memia} propose meMIA, a stacked ensemble MIA that combines NN and LSTM embeddings through a meta-model, demonstrating that multi-level adversarial knowledge improves inference accuracy over single-signal baselines across centralised ML settings. However, meMIA is evaluated in a centralised, shadow-model-dependent setting, and its authors acknowledge that DP remains an open challenge — noting that DP 'poses the same problem to our attack method as all SOTA MI attack methods' - without empirically characterising the extent of that degradation. 

Recent work shows that adversaries combining multiple weak signals like confidence, loss, and model outputs can outperform single-signal baselines~\cite{reza,yeom,Salem2019-qc,carlini2022}. However, a gap remains in evaluating ensemble-based, meta-learning adversaries against strictly DP-bounded models in FL settings~\cite{he2024dpfl,li2025adpf}. Furthermore, existing evaluations predominantly rely on datasets partitioned by the authors themselves, lacking the controls of an independent, third-party benchmark. 

This work addresses these gaps. Unlike meMIA, which operates in a centralised, shadow-model-dependent setting and leaves DP as an open question, our approach extends the ensemble MIA paradigm to FL under an untrusted central aggregator threat model, substituting the shadow-model requirement with auxiliary-record-driven meta-learning. Unlike recent attacks that rely on active data poisoning~\cite{he2024}, image-specific generative models~\cite{shi2026}, or synthetic datasets~\cite{pathade2025}, our approach isolates residual membership signals without adversarial interference or generative assumptions — operating on high-dimensional, tabular genomic data under strictly bounded DP constraints. Evaluation on an independent, third-party benchmark mitigates author-induced partitioning bias. Following the methodology of Carlini et al.~\cite{carlini2022}, the results provide an empirical characterisation of how stacking-based inference degrades across three calibrated DP tiers.

Table~\ref{tab:positioning} summarises how this work relates to prior MIA studies across seven dimensions that collectively define the gap addressed in this paper. \textit{FL Setting} indicates whether the attack was evaluated in a federated learning environment rather than a centralised one. \textit{Shadow-Model-Free} denotes whether the attack 
avoids training shadow models on data drawn from the target's distribution. \textit{Multi-Signal Ensemble} refers to attacks that aggregate diverse inference signals through a learned meta-model, as opposed to thresholding a single metric. \textit{Calibrated DP 
Evaluation} requires testing against multiple, explicitly specified $\epsilon$ tiers rather than a single DP configuration or no DP at all. The remaining attributes --- black-box access, TPR at low FPR reporting, and use of a third-party benchmark --- follow their standard definitions.

\begin{table}[htb]
\centering
\caption{Positioning of this work relative to 
prior MIA studies. \CIRCLE\ = fully addressed, 
\LEFTcircle\ = partially addressed, 
\Circle\ = not addressed.}
\label{tab:positioning}
\resizebox{\linewidth}{!}{%
\begin{tabular}{@{}r c c c c c c c@{}}
\toprule
\textbf{} 
  & \rotatebox{70}{\textbf{FL Setting}} 
  & \rotatebox{70}{\textbf{Black-Box Only}} 
  & \rotatebox{70}{\textbf{Shadow-Model-Free}} 
  & \rotatebox{70}{\textbf{Multi-Signal Ensemble}} 
  & \rotatebox{70}{\textbf{Calibrated DP Evaluation}} 
  & \rotatebox{70}{\textbf{TPR at Low FPR}} 
  & \rotatebox{70}{\textbf{Third-Party Benchmark}} \\
\midrule
Shokri et al.~\cite{reza}
  & \Circle & \CIRCLE & \Circle 
  & \Circle & \Circle & \Circle & \Circle \\
Yeom et al.~\cite{yeom}           
  & \Circle & \CIRCLE & \CIRCLE 
  & \Circle & \Circle & \Circle & \Circle \\
Salem et al.~\cite{Salem2019-qc}  
  & \Circle & \CIRCLE & \CIRCLE 
  & \Circle & \Circle & \Circle & \Circle \\
Nasr et al.~\cite{nasr}           
  & \CIRCLE & \Circle & \CIRCLE 
  & \Circle & \Circle & \Circle & \Circle \\
Jayaraman \& Evans~\cite{jayaraman}  
  & \Circle & \CIRCLE & \LEFTcircle 
  & \Circle & \LEFTcircle & \Circle & \Circle \\
Carlini et al.~\cite{carlini2022} 
  & \Circle & \LEFTcircle & \Circle 
  & \Circle & \LEFTcircle & \CIRCLE & \Circle \\
Ullah et al.~\cite{memia}         
  & \Circle & \CIRCLE & \Circle 
  & \CIRCLE & \Circle & \Circle & \Circle \\
He et al.~\cite{he2024}           
  & \CIRCLE & \Circle & \CIRCLE 
  & \Circle & \Circle & \Circle & \Circle \\
\midrule
\textbf{This work}                
  & \CIRCLE & \CIRCLE & \CIRCLE 
  & \CIRCLE & \CIRCLE & \CIRCLE & \CIRCLE \\
\bottomrule
\end{tabular}%
}
\end{table}

\section{Conclusion}\label{sec:conclusion}

This paper presented an empirical evaluation of Membership Inference Attacks (MIAs) against Privacy-Preserving Federated Learning (PPFL) systems, leveraging the environment of the 2025 NIST Genomics Red Teaming Event. To maximize inference capabilities against an untrusted central aggregator, we introduced an ensemble attack strategy. We combined seven black-box estimators with a meta-classifier, training it to exploit two complementary signals: prediction probabilities from each base attack model and the cross-entropy loss of the target client model. 

Our empirical results indicate a substantial threat posed by ensemble-based adversaries in federated settings. Evaluated across three privacy tiers, an unprotected CNN baseline ($\epsilon=\infty$), a low-DP model ($\epsilon=200$), and a high-DP model ($\epsilon=10$), our methodology achieved the highest score under No DP and Low Privacy. Under High Privacy ($\epsilon=10$), performance converges with other baselines as strong DP noise suppresses all methods toward the random assignment floor. The TPR analysis reveals that even at $\epsilon=200$, the ensemble identifies members at a 30.77\% true-positive rate under a 1\% false-positive budget, demonstrating that average-case accuracy understates the residual risk posed by multi-signal adversary. This confirms that adding calibrated noise smooths the confidence disparities that MIAs traditionally exploit.

These results point to an important implication for the deployment of PPFL in sensitive domains such as genomics. Strict DP constraints (e.g., $\epsilon=10$) substantially mitigate attack success but come at significant utility cost under realistic FL conditions; conversely, a low DP ($\epsilon=200$) preserves utility but does not fully eliminate residual leakage against a multi-signal adversary. Because excessive noise degrades the utility of local model updates, DP alone may be insufficient against multi-signal inference strategies. 

Future work should explore complementary defences such as secure aggregation or output perturbation that can resist meta-classifier attacks without unduly degrading model utility.

\subsubsection*{Acknowledgments}
The author thanks Gary Howarth and the NIST PPFL Genomics Red Teaming team for organising this controlled red-teaming evaluation. The author used Claude Sonnet 4.6 as a writing assistant for grammar, style, and structural feedback. All AI-generated suggestions were reviewed and validated by the author, who retains full responsibility for the content.

\subsubsection*{Disclosure of Interests}
The author has no competing interests to declare that are relevant to the content of this article.

\printbibliography

\end{document}